\documentclass{PoS}
\usepackage{cite}
\usepackage{amsmath} 
\usepackage{mathrsfs}
\usepackage{cite}
\title{Multi-pomeron exchange model for $pp$ and $p\bar{p}$ collisions at ultra-high energy}
\ShortTitle{Multi-pomeron exchange model for $pp$ and $p\bar{p}$ collisions at ultra-high energy}
\author{Evgeniia Bodnia\\
        University of California, Berkeley, USA; Saint Petersburg State University, Russia\\
        E-mail: \email{evgeniya.bodnya@cern.ch}}
\author{Denis Derkach\\
        Oxford University, UK\\
        E-mail: \email{denis.derkach@cern.ch}}
\author{Grigory Feofilov\\
        Saint Petersburg State University, Russia\\
        E-mail: \email{feofilov@hiex.phys.spbu.ru}}
\author{\speaker{Vladimir Kovalenko}
\\
        Saint Petersburg State University, Russia\\
        E-mail: \email{nvkinf@rambler.ru}}
\author{Andrey Puchkov\\
        Saint Petersburg State University, Russia\\
        E-mail: \email{putchkov@mail.ru}}
\abstract{
A new variant of the effective pomeron exchange model is proposed for the description of the correlation, observed in $pp$ and $p\bar{p}$ collisions at center-of-mass energy from SPS to LHC,  between mean transverse momentum and charged particles  multiplicity. The model
is based on the Regge-Gribov approach.
Smooth logarithmic growth with the collision energy was established for the parameter $k$, the mean rapidity density of charged particles produced by a single string.  It was  obtained in the model  by the fitting of the available experimental data on charged particles rapidity density in $pp$
and $p\bar{p}$ collisions. The main effect of the model, a gradual onset of string collectivity
with the growth of collision energy, is accounted by a  free parameter $\beta$ that is responsible in an effective way for the string fusion phenomenon.  Another free parameter, $t$, is used to define string tension. We extract parameters $\beta$ and $t$ from the available experimental results on $\langle p_t \rangle$-multiplicity correlation at nucleon collision energy $\sqrt{s}$ from 17~GeV to 7~TeV.
Smooth dependence of both $\beta$ and $t$ on energy allows to make predictions for the correlation behavior at
the collision energy of 14 TeV. The indications to the string interaction effects in high multiplicity events in $pp$ collisions at  the LHC energies are also discussed.
}

\FullConference{The XXI International Workshop High Energy Physics and Quantum Field Theory,\\
		June 23 -- June 30, 2013 \\
		Saint Petersburg Area, Russia}

\begin{document}

\section{Introduction}
The correlations between mean transverse momentum and charged multiplicity 
(${\langle p_t\rangle \text{-} N_{\text{ch}}}$ -- correlations) 
in hadron collisions have been observed and measured at wide range of 
energy from SPS to the LHC \cite{Anticic-ptN,Marzo-ptN,Aivazyan-ptN,Breakstone-ptN,Albajar-ptN,Arnison-ptN,Abe-ptN,Alexopoulos-ptN,Khachatryan-CMS-ptN}. 
A number of  non-trivial effects 
that were
observed, among them the
are general growth and ``flattening'' of charged particles $p_t$ values with the event 
complexity  
and the increase of mean $p_t$ with the collision energy, could be explained  in terms of some collective processes (such as string fusion, color reconnection, flow etc.), that are relevant also in AA interaction.
In this connection the understanding of the $pp$ collisions at the LHC is important as a base for the analysis of heavy-ion results.

In the present work we develop further the extended multi-pomeron exchange model proposed earlier in \cite{armesto} for analysis of ${\langle p_t\rangle \text{-} N_{\text{ch}}}$ correlations in $pp$ and $p\bar{p}$ collisions.  The model is based on the Regge-Gribov approach, with an effective introduction of the string collectivity effects, performed
in a way, similar to  \cite{armesto}.  But, contrary to  \cite{armesto}, we manage to reduce the number of actually free parameters to only two, that are relevant to string collectivity and string tension. This is achieved, in particular,  by  establishing  a smooth logarithmic growth with the collision energy of the mean rapidity density of charged particles produced by a single string. Below we present the  details of the extended multi-pomeron exchange model followed by the new analysis of available data on
${\langle p_t\rangle \text{-} N_{\text{ch}}}$ correlations in $pp$ and $p\bar{p}$  collisions.

\section{Extended effective multi-pomeron exchange model}

In constructing of the model for $pp$ or $p\bar{p}$ collisions we start from the probability of the event with $n$ cut-pomeron exchanges \cite{Kaidalov,Kaidalov1}:
\begin{equation}\label{prob-w_n}
w_n=\sigma_n / \sum\limits_{n'} \sigma_{n'},
\end{equation}
where $\sigma_n$ is the cross section of $n$ cut-pomeron exchange:
\begin{equation}
\sigma_n=\frac{\sigma_P}{n z} \left( 1-e^{-z}\sum\limits_{l=0}^{n-1} \frac{z^l}{l!}\right), \hspace{1cm} z=\frac{2C\gamma s^\Delta}{R^2+\alpha'\log(s)}
\end{equation}

The numerical values of the parameters used are the following \cite{smth}:
\begin{equation}\nonumber
\Delta = 0.139,\  \alpha'=0.21 \text{ GeV}^{-2},\ 
 \gamma=1.77 \text{ GeV}^{-2},\ 
  R_0^2=3.18 \text{ GeV}^{-2},\  C=1.5\ .
\end{equation}

\begin{figure}
\begin{center}
\includegraphics[width=.4\textwidth]{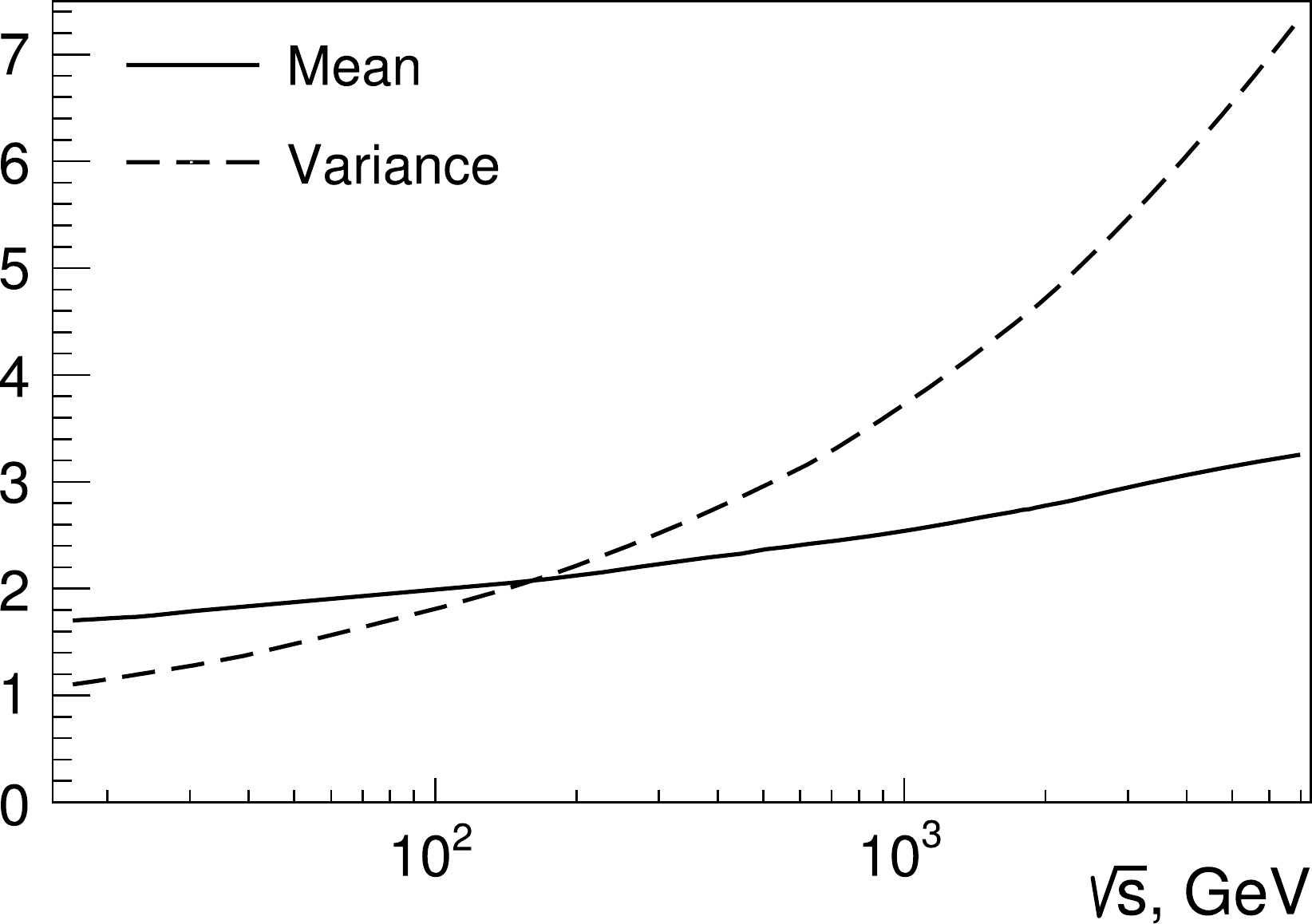}
\end{center}
\vspace{-0.5cm}
\caption{Mean (solid line) and variance (dashed line) of the number of pomerons.}
\label{fig1}
\vspace{-0.5cm}
\end{figure}

Using probabilities (\ref{prob-w_n}) we can calculate
mean and variance of the number of cut-pomerons:
\begin{equation}\nonumber
\langle n \rangle = \sum\limits_{n'=0}^{\infty} n' \ w_{n'}, \  \text{Var }n = \sum\limits_{n'=0}^{\infty} {n'}^2 \ w_{n'} - \langle n \rangle^2 \ .
\end{equation}
The results are plotted in Fig. \ref{fig1} as a function of collision energy $\sqrt{s}$.

For the description of the charged multiplicity we assume that each cut-pomeron corresponds to a pair of strings, 
each of them emits particles into given acceptance according to Poisson
distribution with mean value, proportional 
to this (pseudo)-rapidity interval.
Therefore the probability for $n$ cut-pomerons to
produce $N_{\text{ch}}$ particles is given by:

\begin{equation}\label{prob-one-pomeron}
P(n,N_{\text{ch}})=\text{exp}(-2nk\delta)\frac{(2nk\delta)^{N_{\text{ch}}}}{N_{\text{ch}}!},
\end{equation}
where $k$ is the mean multiplicity per rapidity unit from one string, $\delta$ is the acceptance i.e. width of \mbox{(pseudo-)rapidity} interval.

Using formula (\ref{prob-one-pomeron}) we obtain the
probability to have $N_{\text{ch}}$ particles in a given event:
\begin{equation}\nonumber
  \mathscr{P}(N_{\text{ch}})=\sum\limits_{n=1}^\infty
  w_n P(n,N_{\text{ch}})
\end{equation}
and mean charged multiplicity:
\begin{equation}\label{Nch-formula}
\langle N_{\text{ch}} \rangle =
 \sum\limits_{N_{\text{ch}}=0}^\infty N_{\text{ch}} \mathscr{P}(N_{\text{ch}}).
\end{equation}

For the description of pseudorapidity 
density of charged particles we
can use formula (\ref{Nch-formula}), substituting
$\delta=1$. Note, that in this case we have only one parameter $k$ to adjust for the description of 
experimental data.

In order to cover also 
the transverse momentum and its
correlation with multiplicity, the extension of the model is performed as follows.
We use the general idea of the Schwinger mechanism of particles production from one string \cite{Schwinger}, 
that the transverse momentum distribution
of charged particles from a string has a  Gaussian
form:

\begin{equation}\label{Schwinger}
\frac{d^2 N_{\text{ch}}} {dp_{t}^2} \sim
\text{exp} \left( \frac{-\pi(p^2_t+m^2)}{t} \right),
\end{equation}
where $t$ corresponds to the string tension.

The assumptions that all strings are identical or that the string tension does not depend on multiplicity
would give zero correlation between transverse 
momentum and multiplicity.
In order to include collectivity in our model and to be able to describe non-trivial $\langle p_t \rangle\text{-} N_{\text{ch}}$ correlations we
introduce, similar to \cite{armesto}, the new parameter $\beta$, which is responsible in an effective way for string interaction (fusion), with $\beta=0$ corresponding to the absence of collective effects. Thus the dependence of transverse momentum from
one string on the number of cut-pomerons is introduced:
\begin{equation}\label{Schwinger-mod}
\frac{d^2 N_{\text{ch}}} {dp_{t}^2} \sim
\text{exp} \left( \frac{-\pi(p^2_t+m^2)}{n^\beta t} \right),
\end{equation}
Note that denominator in (\ref{Schwinger-mod}) gives effective 
string tension $n^\beta t$.

$\langle p_t \rangle\text{-} N_{\text{ch}}$ correlation function in the present model is calculated as

\begin{equation}\label{pt-Nch-function}
   \langle p_{t} \rangle_{N_{\text{ch}}}=\dfrac{\int\limits_0^\infty
   {\rho(N_{\text{ch}},p_t)p_t^2 dp_{t}}} {\int\limits_0^\infty
   {\rho(N_{\text{ch}},p_t)p_t dp_t}},
\end{equation}
where $\rho(N_{\text{ch}},p_t)$ is the distribution of  $N_{ch}$ particles
over $p_t$:
\begin{equation}\label{rho-distribution}
\rho(N_{\text{ch}},p_t)=\dfrac{C_w}{z} 
\sum\limits_{n=1}^{\infty} \dfrac{1}{n}\left( 1- \exp(-z)\sum\limits_{l=0}^{n-1}\frac{z^l}{l!}\right) \cdot
\text{exp}(-2nk\delta)\dfrac{(2nk\delta)^{N_{\text{ch}}}}{N_{\text{ch}!}}\cdot
\dfrac{1}{n^\beta t} \text{exp} \left(-\dfrac{\pi p_t^2}{n^\beta t} \right).
\end{equation}
Here $C_w$ is a normalization factor.

The first multiplier in (\ref{rho-distribution}) corresponds to the
probability of the production of $n$ pomerons, the second one is the 
Poisson distribution of the charged particles from $2n$ strings, 
and the last reflects the modified Schwinger mechanism,
discussed above.

\section{Parameters determination and results}
\begin{figure}[h]
\begin{minipage}[h]{0.475\linewidth}
\center{\includegraphics[height=0.7\textwidth]{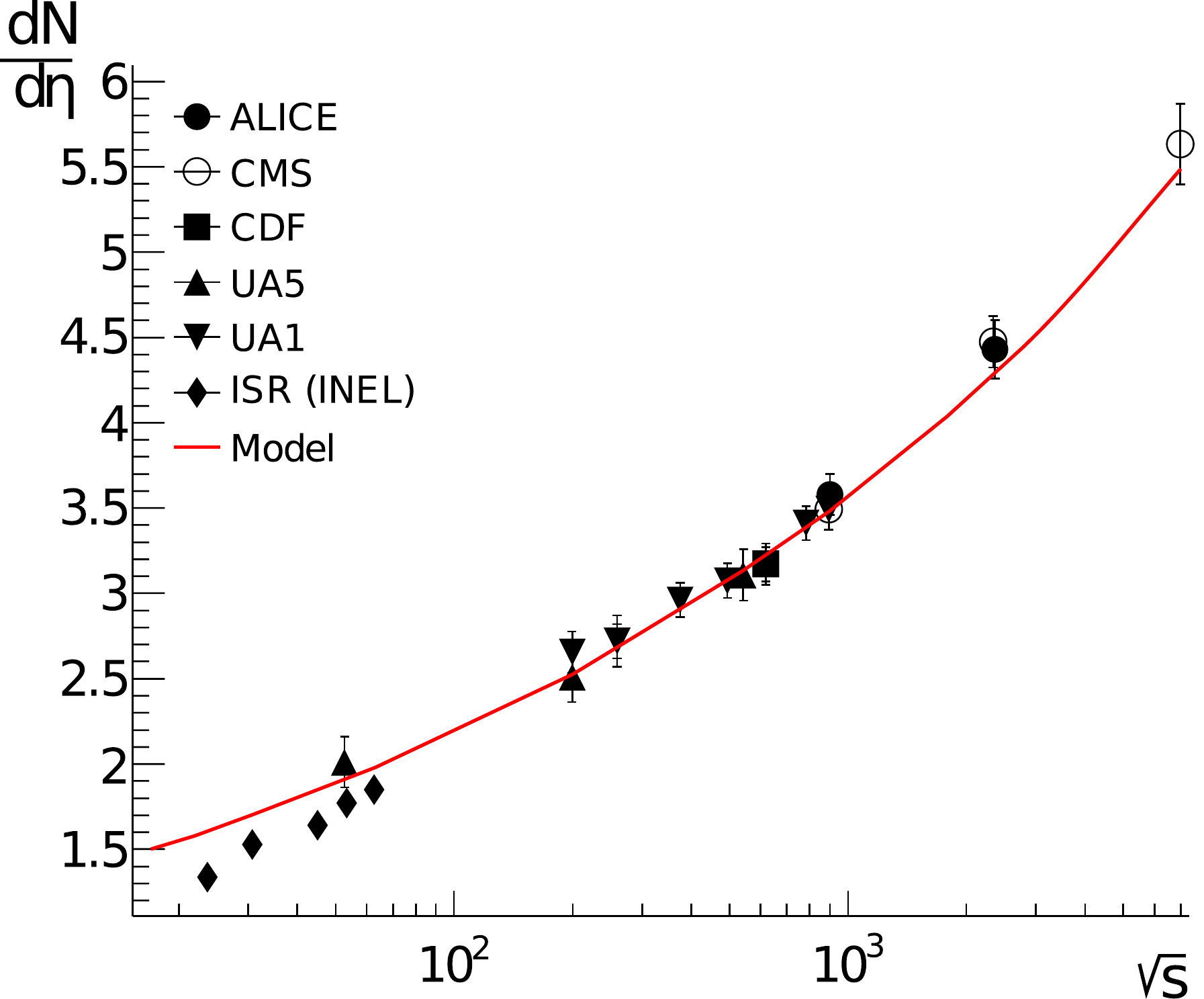}} 
\caption{Pseudorapidity density of multiplicity as a function on energy.
Experimental data are borrowed from \cite{ALICEpA}.}
\label{fig2}
\end{minipage}
\hfill
\begin{minipage}[h]{0.475\linewidth}
\center{\includegraphics[height=0.7\textwidth]{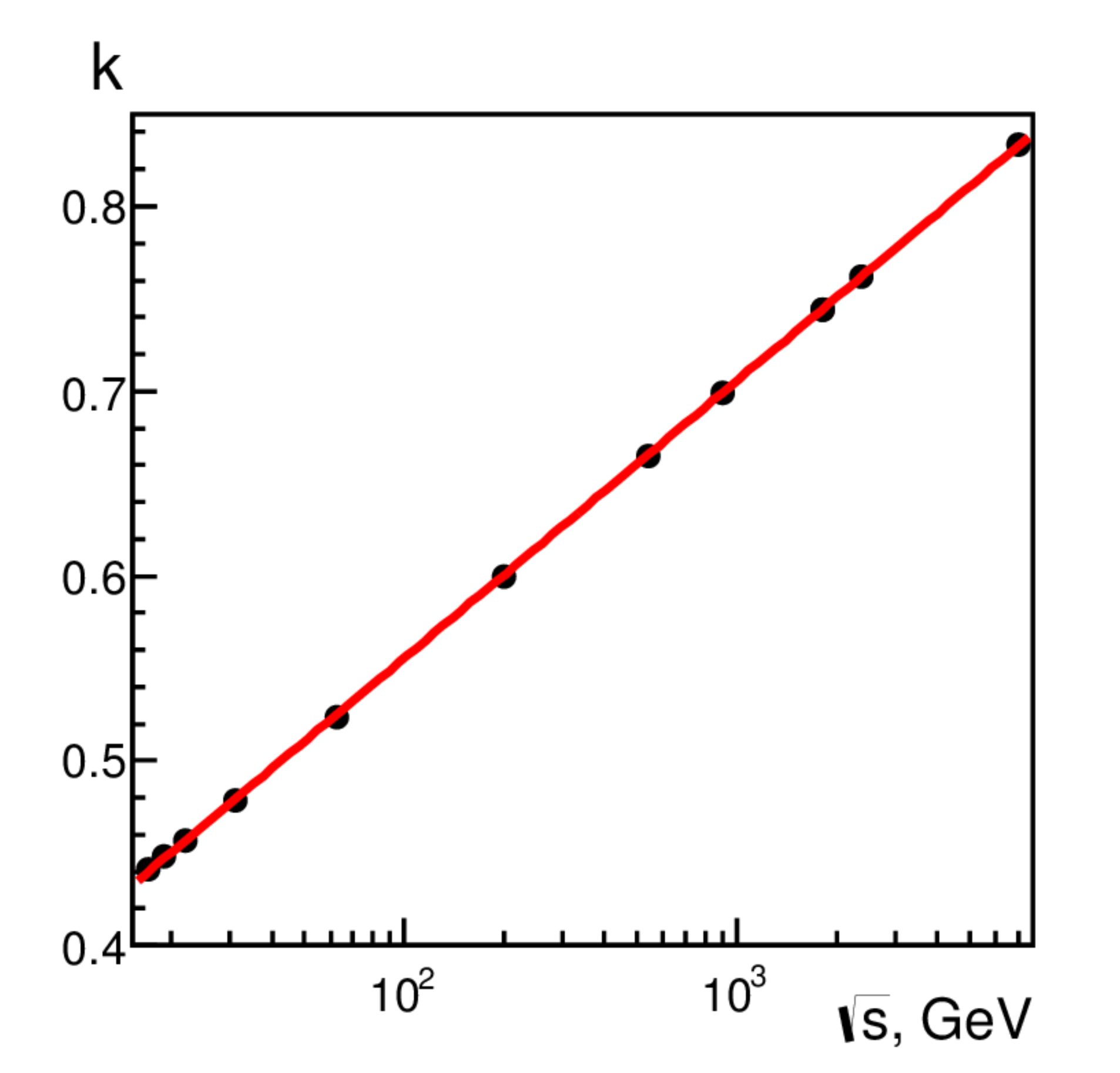}} 
\caption{Dependence of parameter $k$ on energy,
obtained in the present model.
\newline
}
\label{fig3}
\end{minipage}
\end{figure}

The first stage of the parameters fixing is the
determination of the parameter $k$ -- mean multiplicity
per rapidity from one string. This parameter is extracted
from the experimental data (Fig.~\ref{fig2}) on charged multiplicity:
\begin{equation}\label{Nch-s}
\langle N_{\text{ch}} \rangle (s) = \sum\limits_{N_{\text{ch}}}^\infty N_{\text{ch}} \mathscr{P}(N_{\text{ch}}) = 2\langle n\rangle\cdot k\cdot \delta.
\end{equation}
Smooth logarithmic growth of parameter $k$ is obtained (Fig. \ref{fig3}): $k=0.255+0.0653 \ln \sqrt{s}$.

\begin{figure}
\hspace{-1.0cm}
\begin{minipage}[h]{1.15\linewidth}
\includegraphics[width=.36\textwidth]{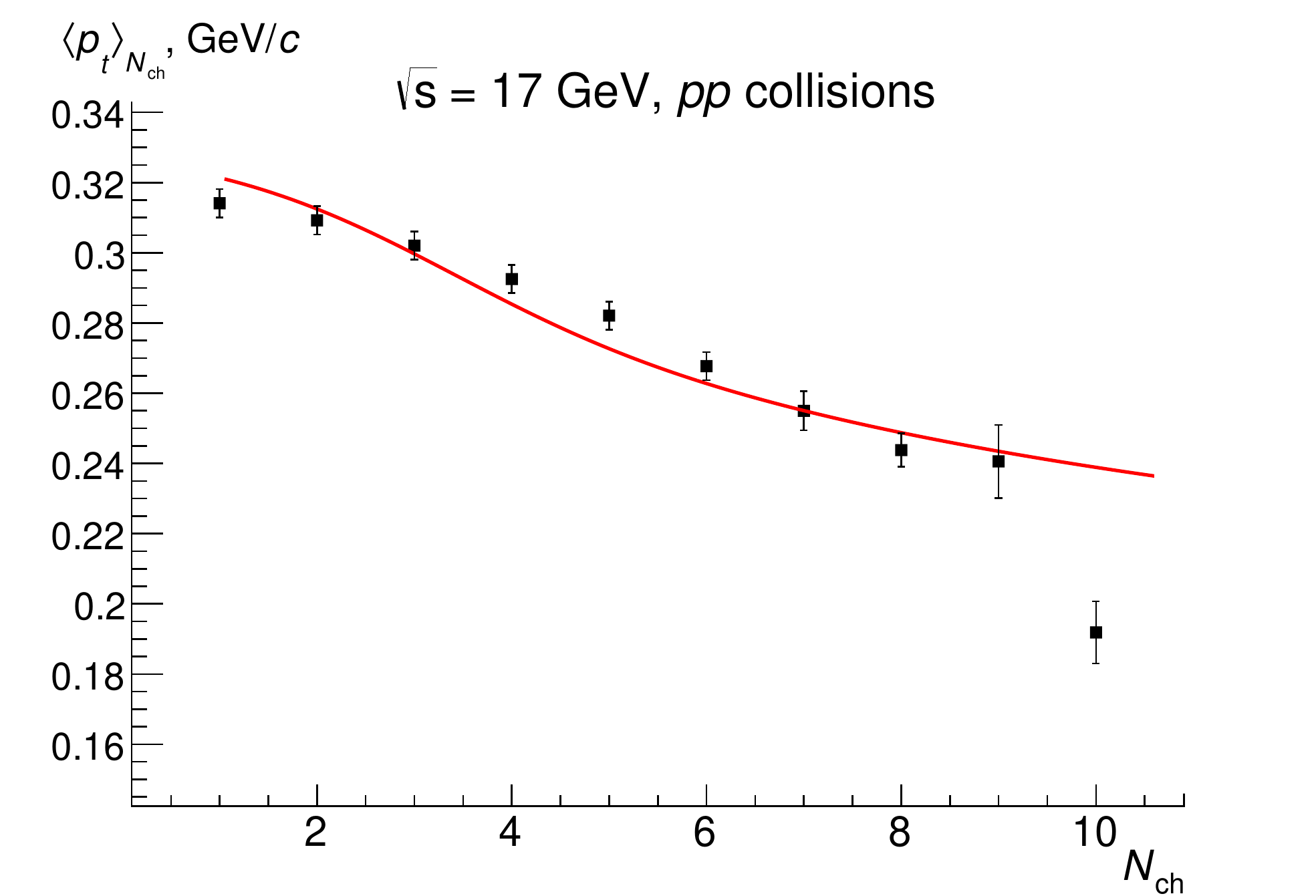}\hspace{-0.7cm}
\includegraphics[width=.36\textwidth]{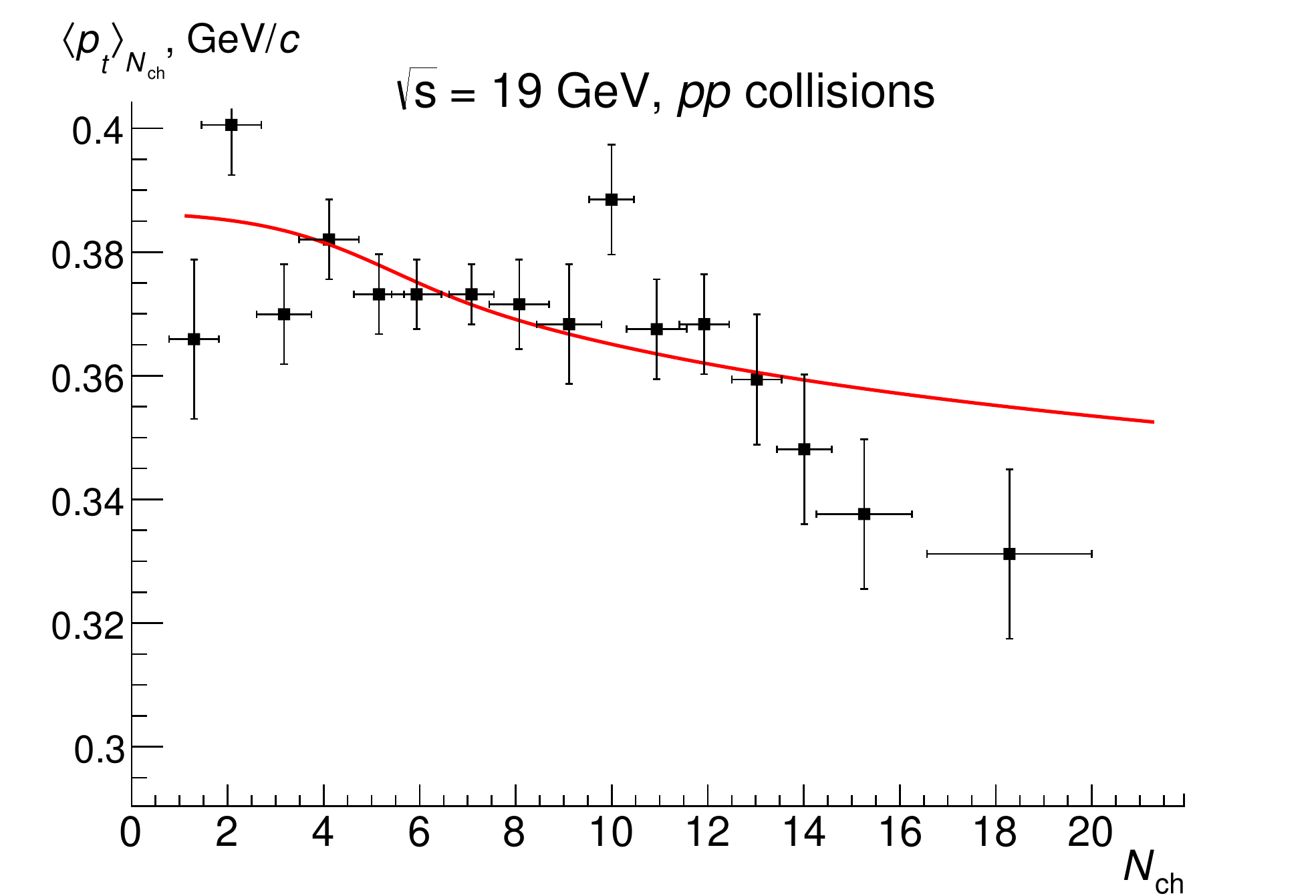}\hspace{-0.7cm}
\includegraphics[width=.36\textwidth]{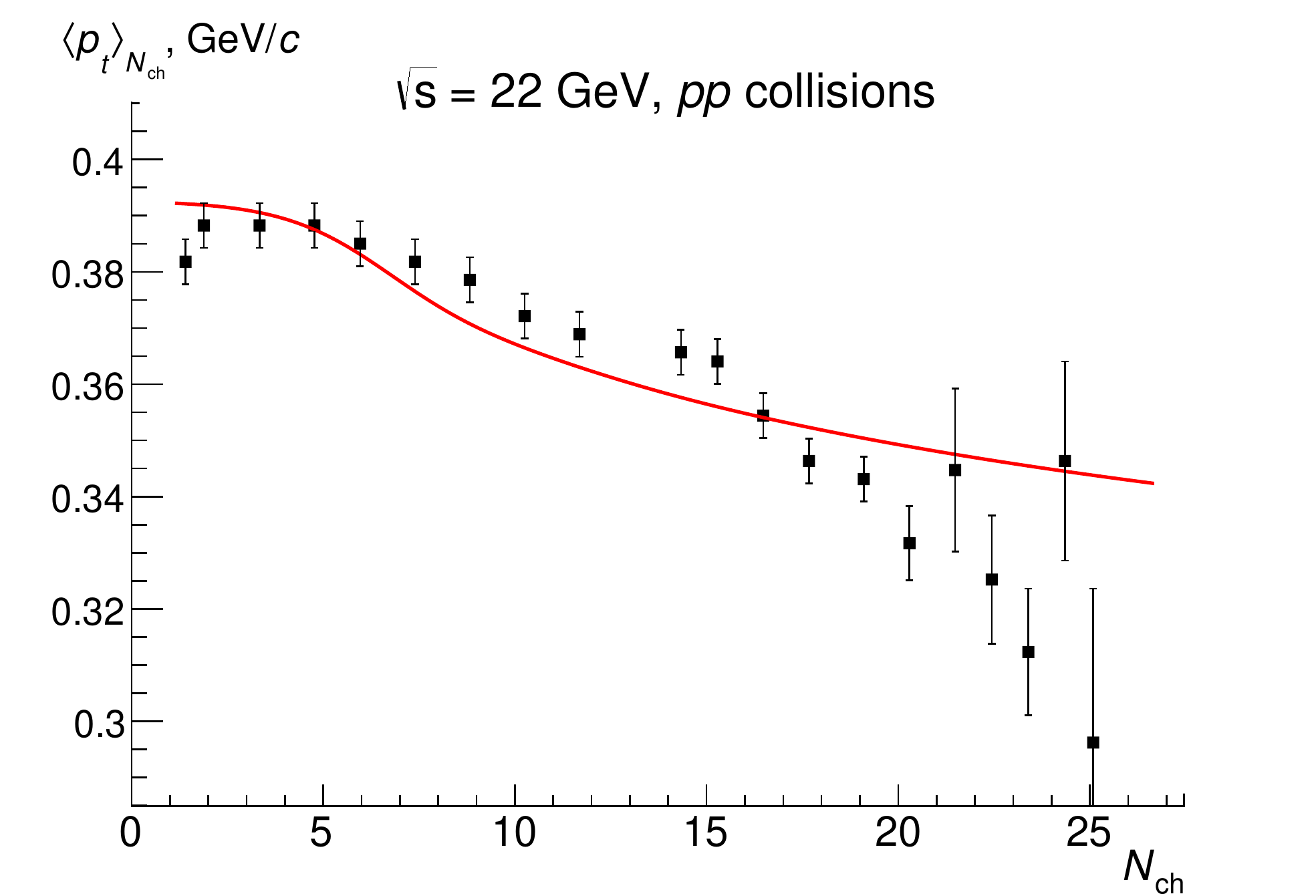}
\\
\\
\includegraphics[width=.36\textwidth]{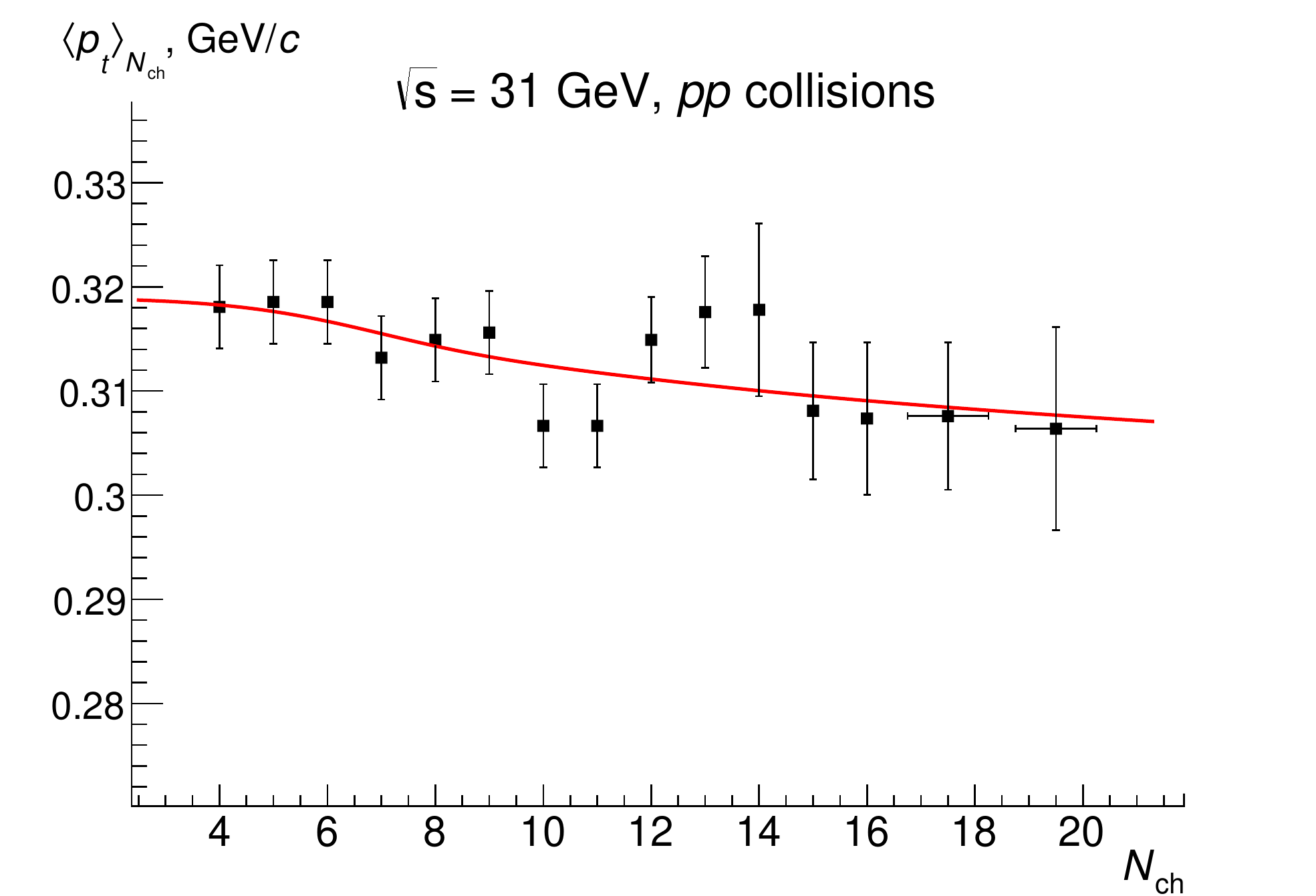}\hspace{-0.7cm}
\includegraphics[width=.36\textwidth]{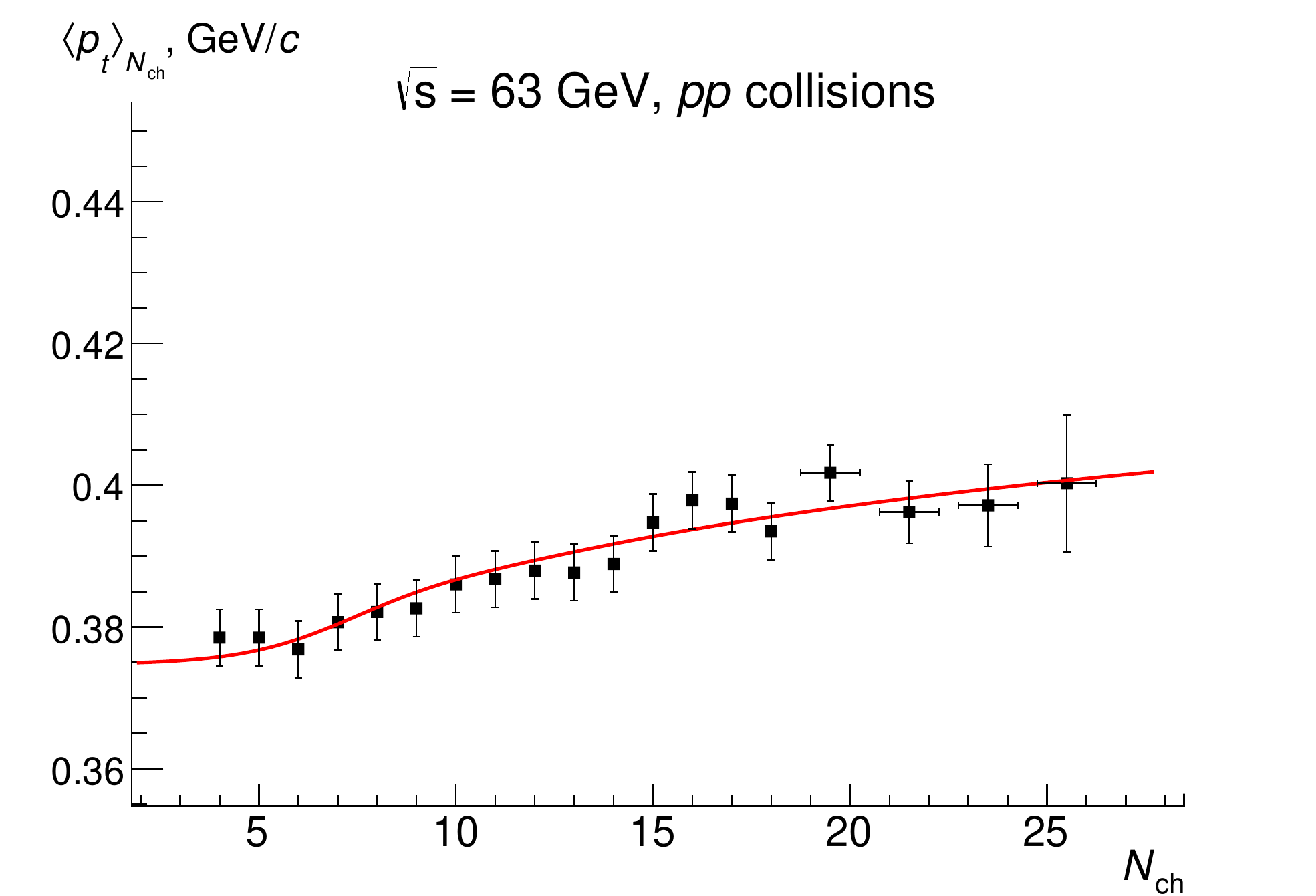}\hspace{-0.7cm}
\includegraphics[width=.36\textwidth]{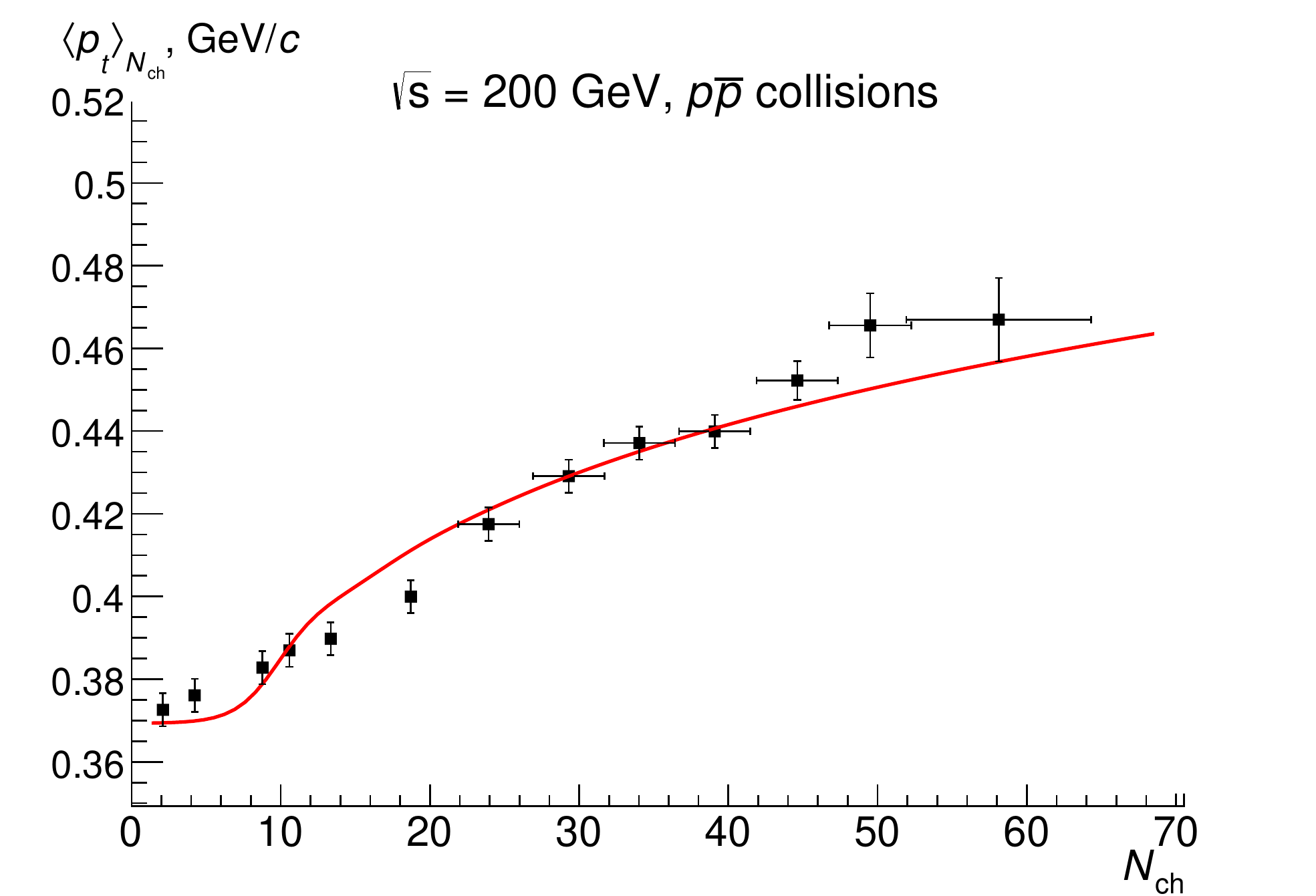}
\\
\\
\includegraphics[width=.36\textwidth]{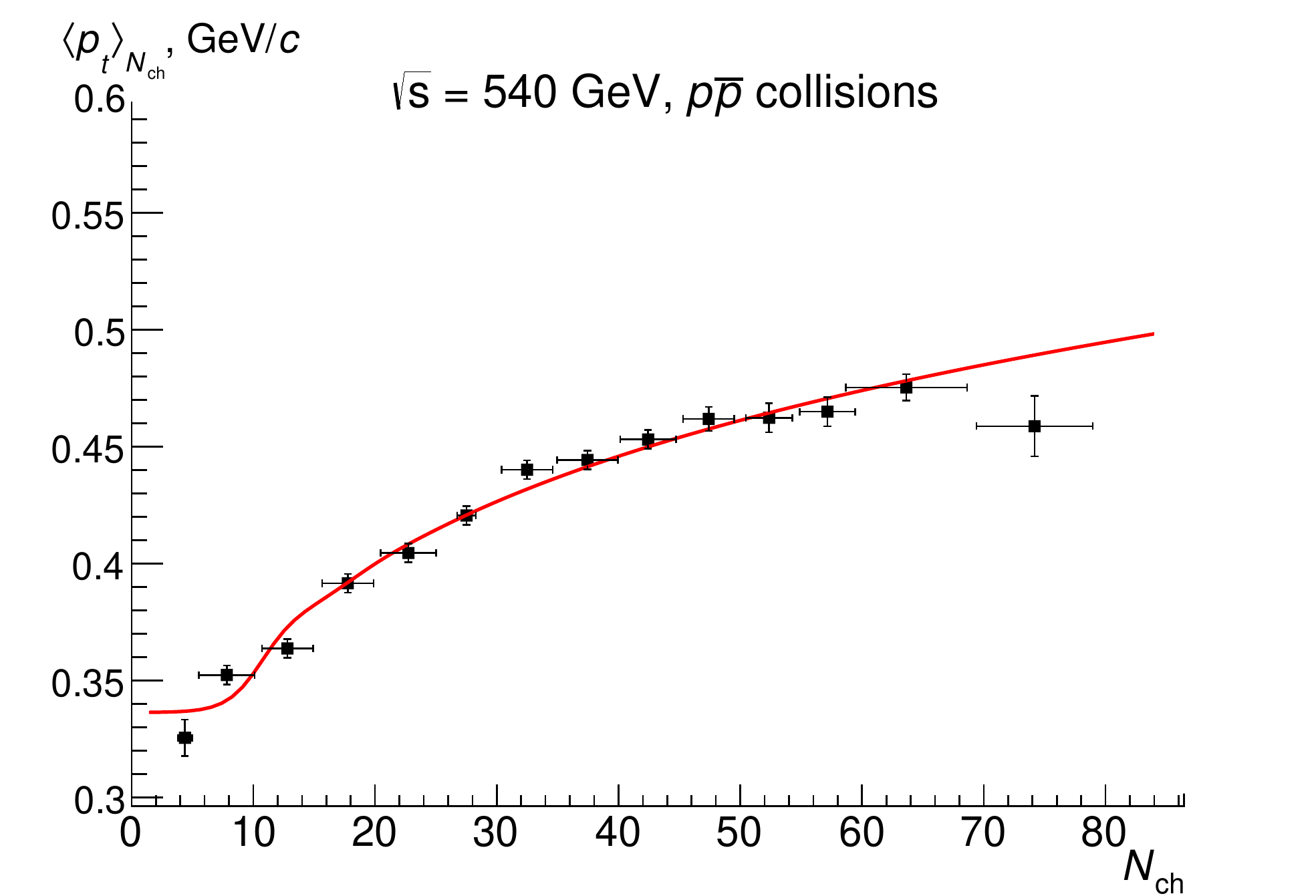}\hspace{-0.7cm}
\includegraphics[width=.36\textwidth]{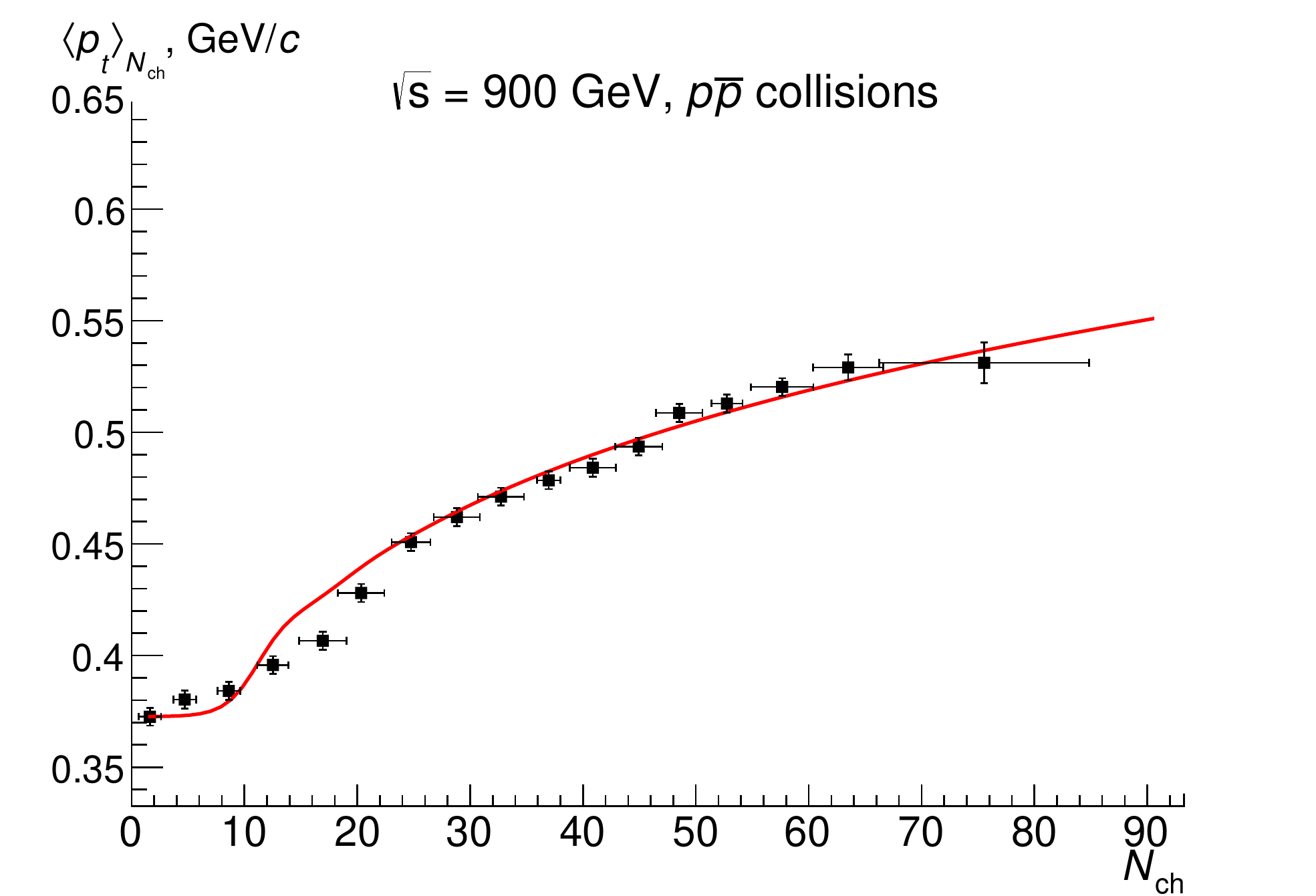}\hspace{-0.7cm}
\includegraphics[width=.36\textwidth]{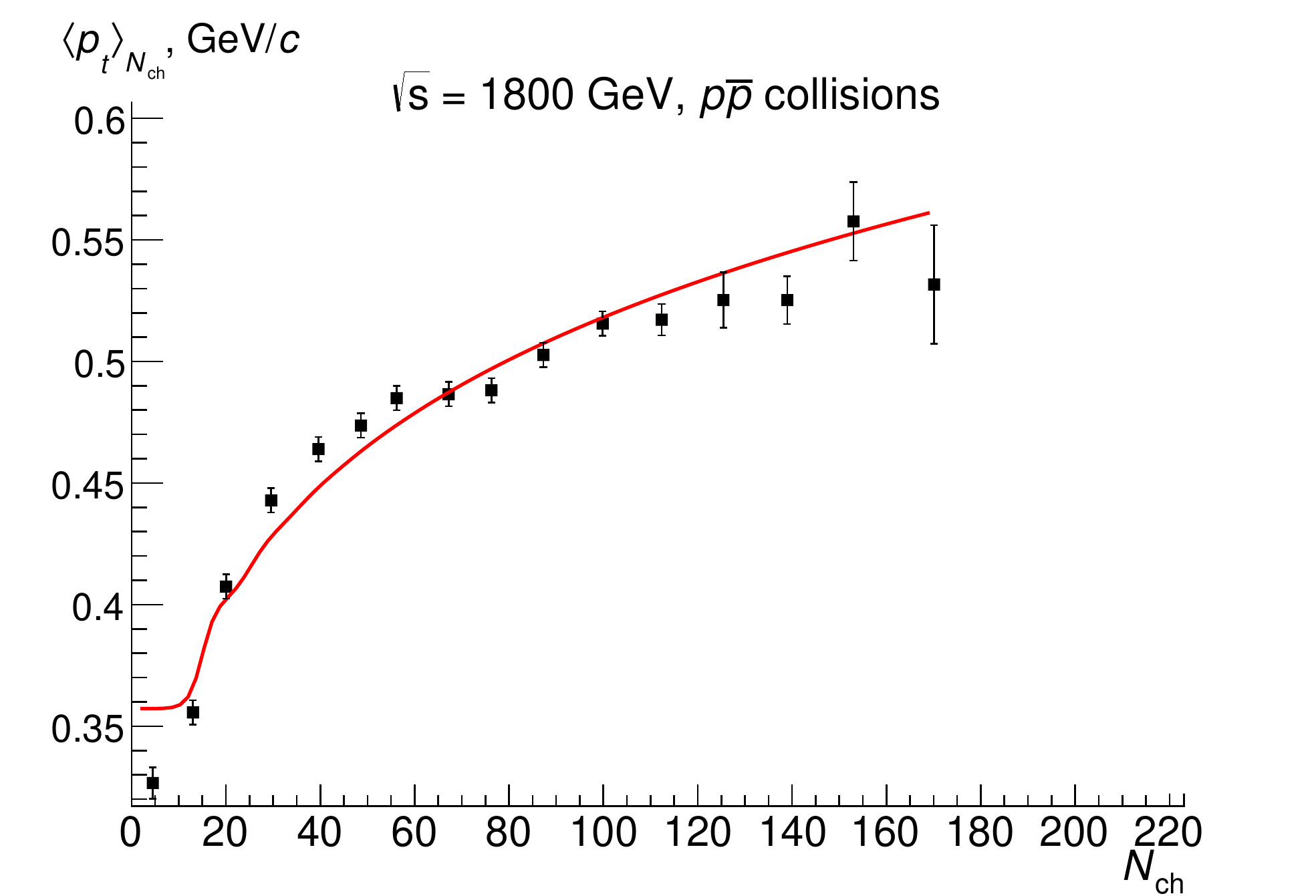}
\\
\\
\includegraphics[width=.36\textwidth]{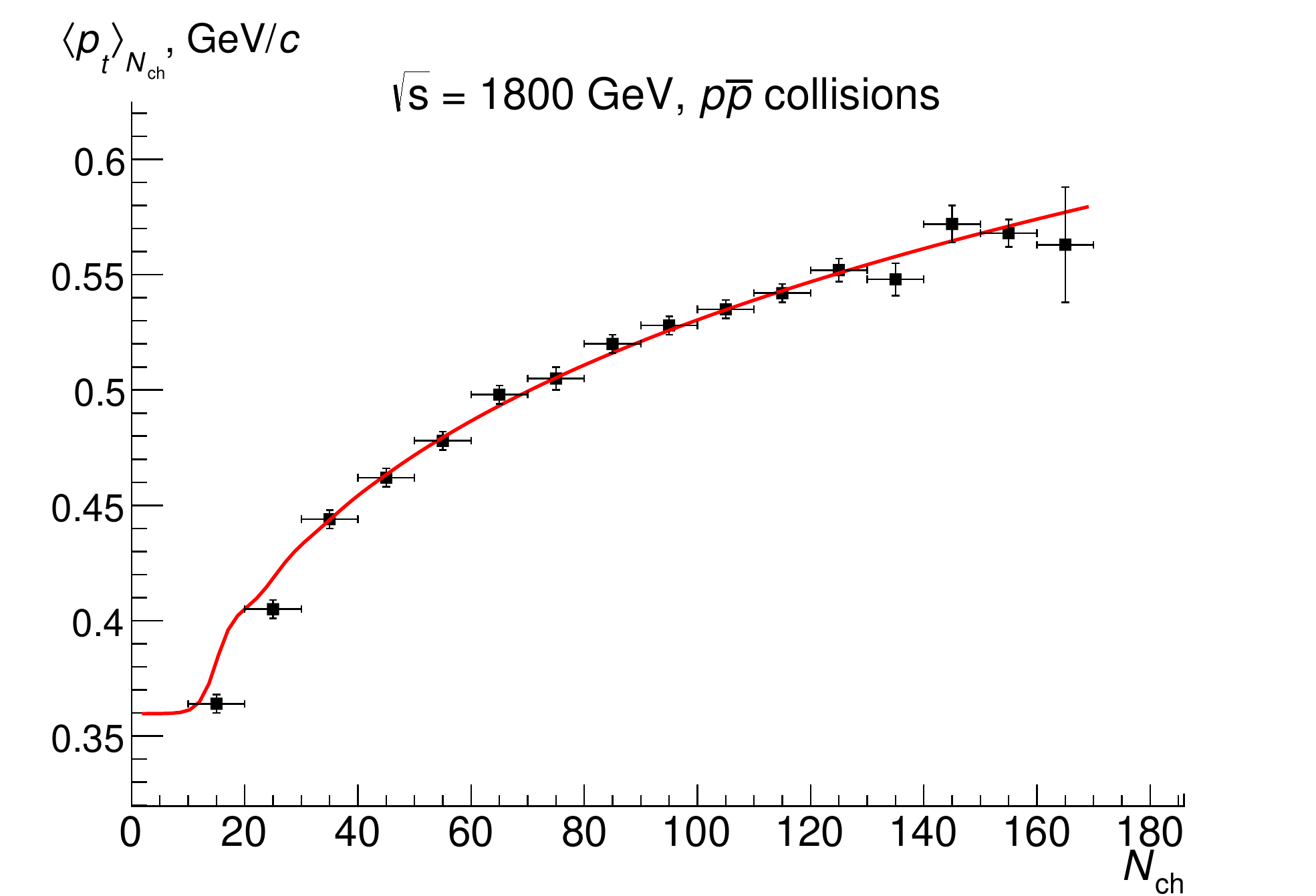}\hspace{-0.7cm}
\includegraphics[width=.36\textwidth]{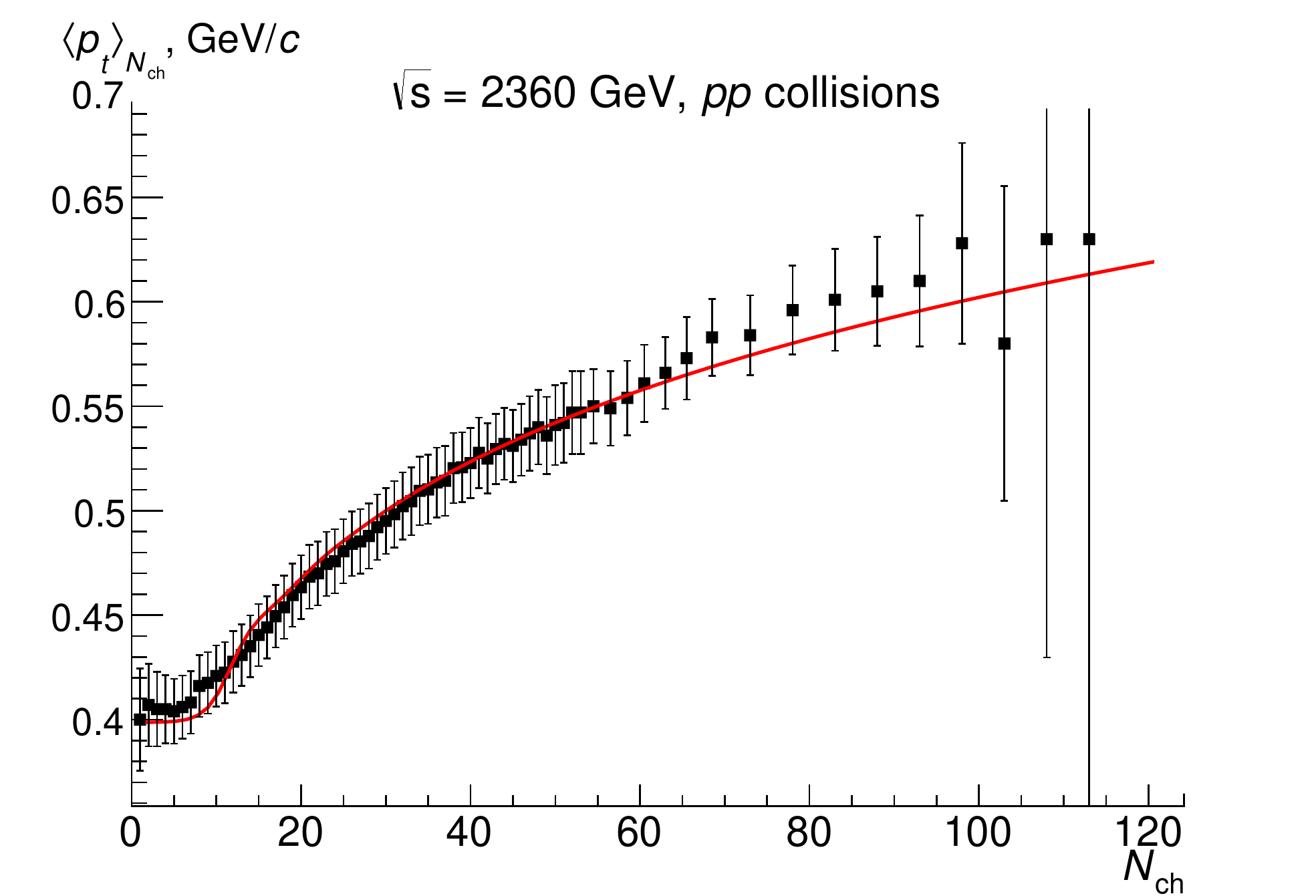}\hspace{-0.7cm}
\includegraphics[width=.36\textwidth]{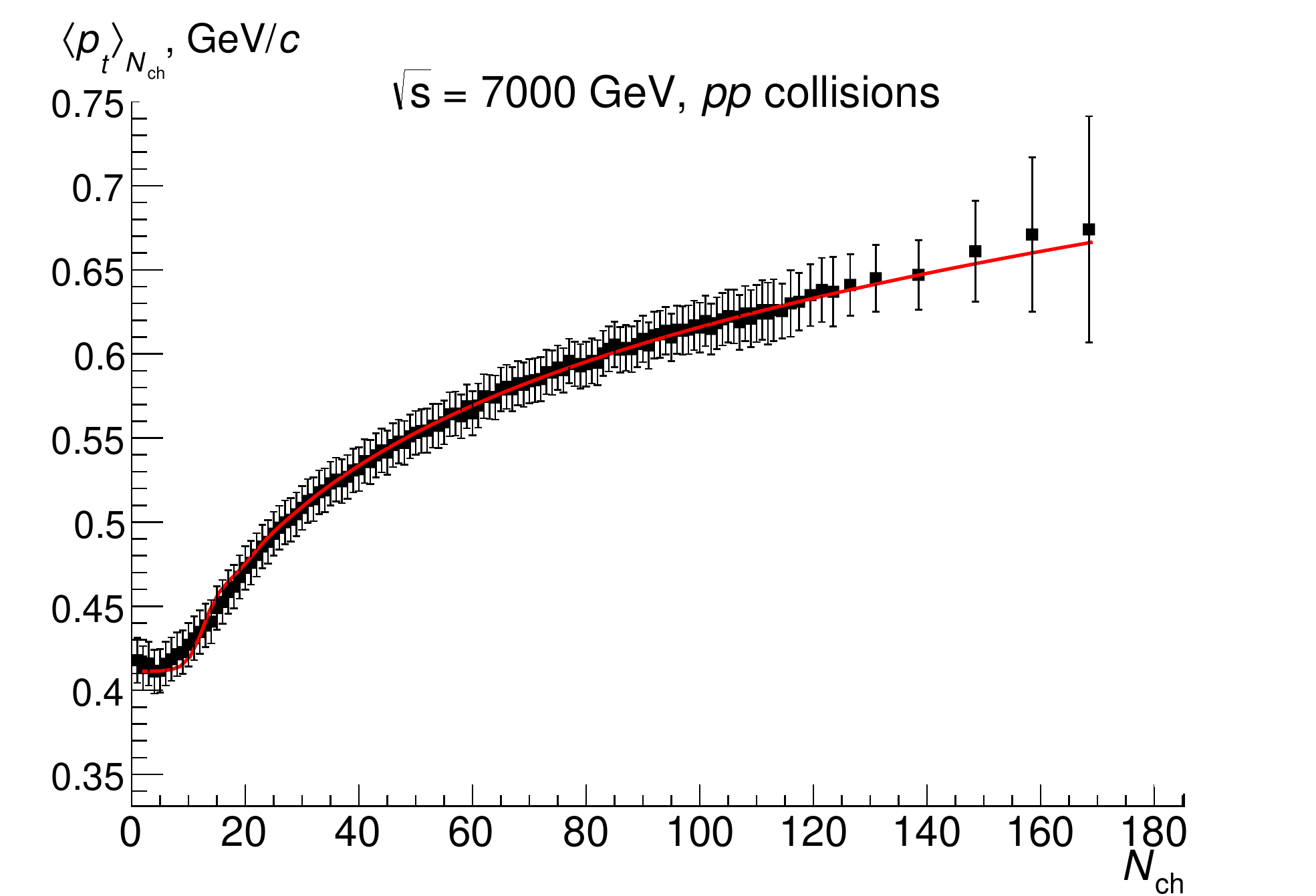}
\end{minipage}

\caption{Experimental data \cite{Anticic-ptN,Marzo-ptN,Aivazyan-ptN,Breakstone-ptN,Albajar-ptN,Arnison-ptN,Abe-ptN,Alexopoulos-ptN,Khachatryan-CMS-ptN}, fitted by the present model.}
\label{fig4}
\end{figure}

\begin{figure}
\begin{center}
\includegraphics[width=.70\textwidth]{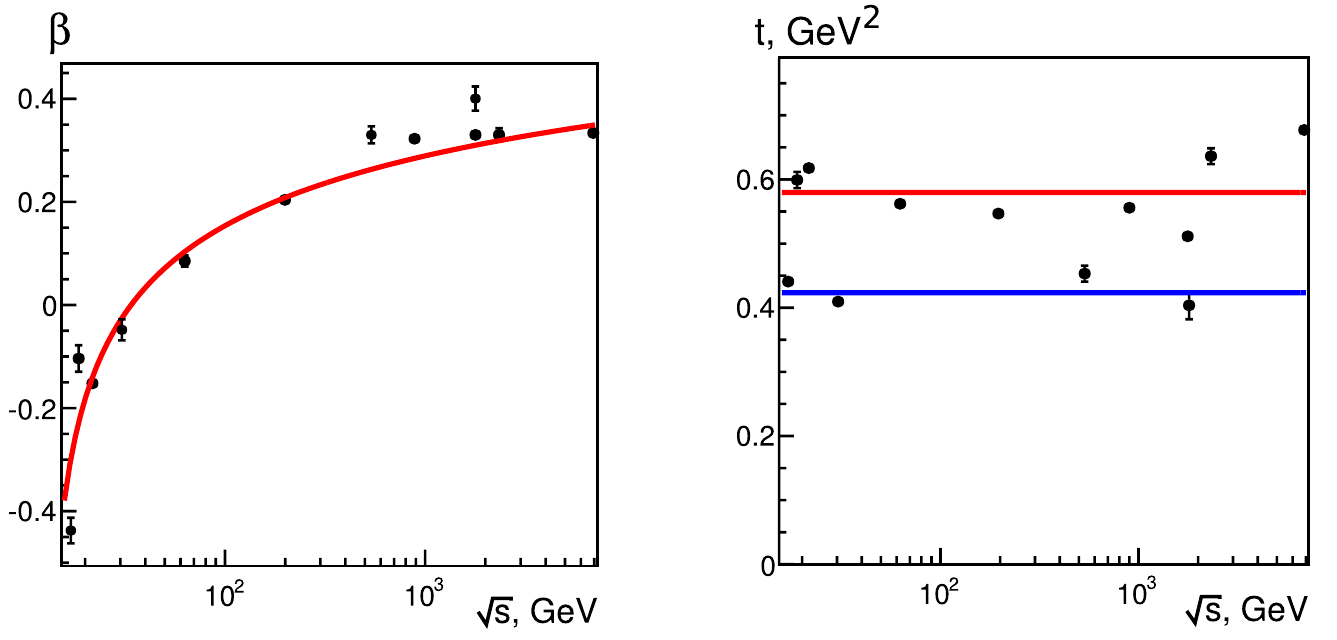} \hspace{1.2cm}
\end{center}
\vspace{-0.7cm}
\caption{Dependence of parameters $\beta$ (left) and $t$ (right) on energy.}
\label{fig5}
\vspace{-0.3cm}
\end{figure}

The values of the parameters $\beta$ and $t$ are obtained using
the data on ${\langle p_t \rangle\text{-} N_{\text{ch}}}$ correlations
 from 17 GeV to 7 TeV. Results
of fitting are shown in Fig. \ref{fig4}.
Parameters $\beta$ and $t$ as function on energy
are shown in Fig. \ref{fig5}. 
The smooth behaviour of parameter $\beta$ with energy is obtained 
and approximated in Fig. \ref{fig5} by 
$\beta=1.16[1-(\ln\sqrt{s}-2.52)^{-0.19 }].$
Similar to \cite{armesto} the set of obtained variables $t$ have split into two subsets:
one is around ${t=0.566\text{ GeV}^2}$ and the second ${t=0.428\text{ GeV}^2}$.
This discrepancy between them may be related to differences in 
data analysis procedures and interpolation to the softest part of $p_t$ spectra,
performed by various experiments.
The points belonging to the first subset are used for the further analysis because they provide the correct values of $\langle p_t \rangle$ (see \cite{armesto}).
\subsection{Predictions for the LHC at 14 TeV.}
Using obtained dependence of the parameters ($k, \beta, t$)
on energy we extrapolated their values up to the energy
$\sqrt{s}=14\text{ TeV}$ and calculated the $\langle p_t \rangle\text{-} N_{\text{ch}}$ correlation function.
For the parameter $t$ we used three assumptions: no dependence on energy
($t=0.566 \text{ GeV}^2$), parabolic fit ($t=0.731 \text{ GeV}^2$)
and $t=0.676 \text{ GeV}^2$ -- the same as at 7 TeV.
The results are shown in Fig. \ref{fig6}. The experimental
data at this energy would enable to make final decision
concerning the behaviour of $t$ with energy.
\subsection{Discussion of the possible relation to the string fusion}
The parameters, obtained after the fitting procedure, demonstrate a growth
of multiplicity from one string ($k$) with increasing energy,
accompanied by an increase of transverse momentum. This 
result agrees with the basic concept of string fusion model \cite{StringFusion0, StringFusion}, according to which at high energies strings with greater tension are formed due to the overlap in the transverse plane. 
In line with  this model, the average multiplicity per rapidity unit
and mean square of the transverse
\begin{figure}[h]
\begin{minipage}[h]{0.455\linewidth}
\includegraphics[height=0.6\textwidth]{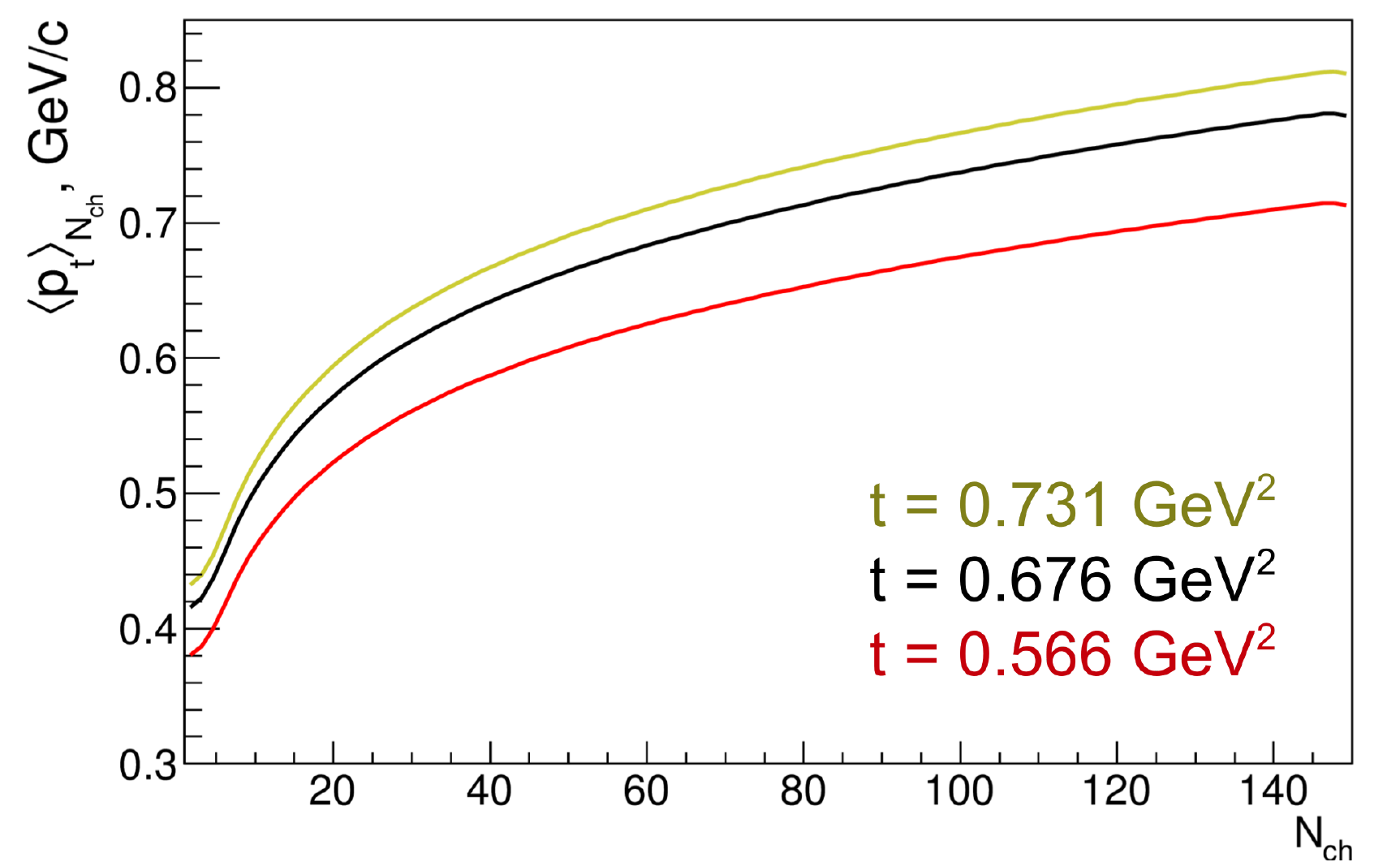}\hspace{0.5cm}
\caption{Predictions for the $\langle p_t\rangle \text{-} N_{\text{ch}}$ 
correlation function at 14 TeV, obtained for
$|\eta|<2.4$\,.}
\label{fig6}
\end{minipage}
\hfill
\begin{minipage}[h]{0.455\linewidth}
\includegraphics[height=0.6\textwidth]{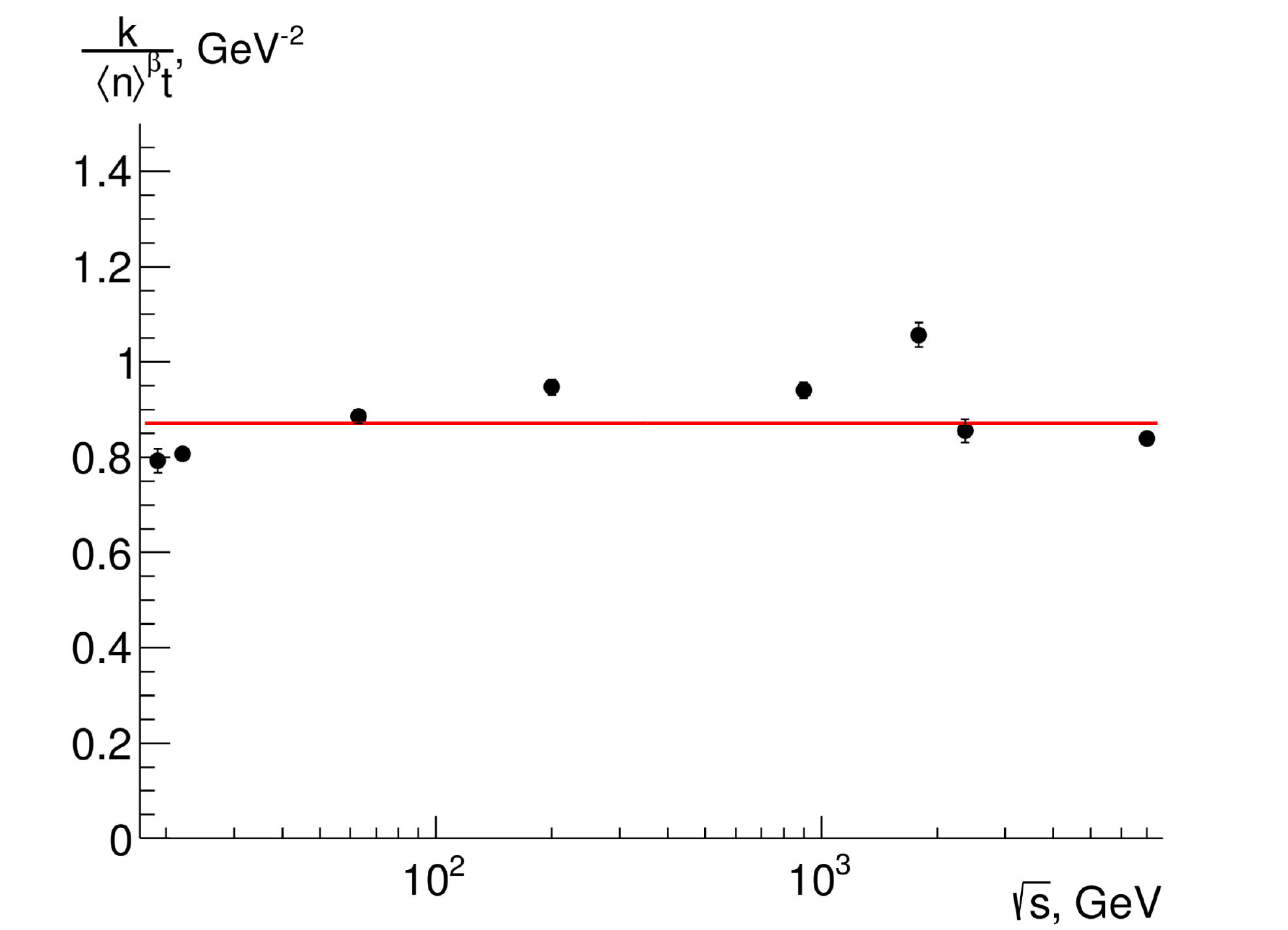}
\newline
\caption{Ratio of mean multiplicity from one source
over characteristic $p_t$ from one source.
}
\label{fig7}
\end{minipage}\vspace{-0.1cm}
\end{figure}
momentum of charged particles from a
cluster of fused strings are proportional to the square root of the number of strings overlapping:
  $
   \mu = \mu_0 \sqrt {\eta}, p_t^2 = p_0^2 \sqrt{\eta}.
  $
Thus, the ratio of these quantities in average should not depend on
energy:
$   \dfrac  {\mu} {p_t ^ 2}=
   \dfrac  {\mu_0} {p_0 ^ 2}\,.$
   
Since in our model the product $ {\langle n \rangle} ^ \beta t $ is
a characteristic square of the transverse momentum of the particles from one
source, 	fulfilment of this condition can be checked
by plotting the ratio $ k / ({\langle n
\rangle} ^ \beta t) $ as a function of energy. 
The results presented in Fig.  \ref {fig7} indeed show no dependence of this ratio on energy.
Thus, we can conclude that the experimental data support the hypothesis of the string fusion as the source of collectivity in $ pp $ and $ p \bar {p} $ collisions in a wide energy range.
\section{Conclusions}\vspace{-0.1cm}
 The extended multi-pomeron exchange  model with collectivity effects is applied to the analysis of experimental data on $\langle p_t\rangle \text{-} N_{\text{ch}}$ correlation in a wide energy range of $ pp $ and $ p \bar {p} $ collisions.
\newline
Thus it was possible to establish:

(i)               Smooth logarithmic growth with the collision energy of mean multiplicity ($k$) of charged particles from one string;

(ii)              Smooth growth of the string collectivity parameter ($\beta$) with the collision energy;

(iii)            Stability of the string  tension parameter ($t$), which is   consistent with a constant.

 Results are pointing at the numerical agreement with string fusion model.
 Predictions for ${\langle p_t\rangle \text{-} N_{\text{ch}}}$ correlations at the future LHC collision energy of
14 TeV were obtained.


\end{document}